\documentclass{appolb}

\def\lQ{\Lambda_{\rm QCD}}

\newcommand{\be}{\begin{equation}}
\newcommand{\ee}{\end{equation}}
\newcommand{\bea}{\begin{eqnarray}}
\newcommand{\eea}{\end{eqnarray}}

\def\als{\alpha_{\rm s}}
\def\siml{{\ \lower-1.2pt\vbox{\hbox{\rlap{$<$}\lower6pt\vbox{\hbox{$\sim$}}}}\ }}
\def\simg{{\ \lower-1.2pt\vbox{\hbox{\rlap{$>$}\lower6pt\vbox{\hbox{$\sim$}}}}\ }}

\begin{document}
\pagestyle{plain}

\newcount\eLiNe\eLiNe=\inputlineno\advance\eLiNe by -1
\title{HEAVY QUARKONIUM PHYSICS\\ --THEORETICAL STATUS
\thanks{
Presented by N. B. at the XXXI Conference 
``Matter to the Deepest'', Ustron, Poland, 5-11 September 2007.
Work  supported by  the  European Research Training Network FLAVIA{\it net} 
(FP6, Marie Curie Programs, Contract MRTN-CT-2006-035482).
}}
\author{Nora Brambilla and Antonio Vairo
\address{Dipartimento di Fisica dell'Universit\'a 
di Milano and INFN, via Celoria 16, 20133 Milano, Italy}}

\maketitle
\begin{abstract}
We briefly review the theoretical status and the open  theoretical challenges
in the physics of heavy quarkonium.
\end{abstract}

\section{Interest of Heavy Quarkonium Physics}
Systems made by two heavy quarks are particularly interesting from the theoretical 
point of view. They are characterized by the energy scales typical of a nonrelativistic bound system:
the scale of the mass $m$, the scale of the  relative momentum $p \sim mv \sim r^{-1}$,
the scale of the binding energy $E \sim m v^2$, $v \ll 1$ being the quark velocity
and $r$  the radius of the system.  
This is similar to what happens for the hydrogen  atom or for positronium  
in QED. The heavy quarks however interact strongly and their bound state dynamics is 
determined by QCD and subjected to confinement \cite{Brambilla:1999ja}. Besides the scales
listed above, one has therefore to consider also $\lQ$, the scale at which nonperturbative 
effects become important. The  specific feature of being  multi-scale makes 
heavy quarkonium an interesting probe for  several  energy regimes of QCD, from the hard
region, where an expansion in the coupling constant $\als$ is legitimate, to
the low energy region, where QCD nonperturbative effects dominate. 
In particular the mass  scale is ``hard'', $m \gg \lQ$,  
and the physics at such scale may be calculated with a perturbative expansion in $\als$.
The relative momentum or ``soft'' scale, proportional to the  inverse size of the system,
may be a perturbative ($\gg \lQ$) or a nonperturbative scale ($\sim \lQ$) depending 
on the physical systems. Finally, only for $t\bar{t}$ threshold states  
the binding energy, i.e. the  ``ultrasoft'' scale,  may still be perturbative. 
Heavy quark-antiquark states are thus an ideal and to some extent unique 
laboratory where our understanding of nonperturbative QCD, its interplay with perturbative 
QCD and the behaviour of the perturbative series in the bound state may be tested 
and understood in a controlled framework. This has been so historically, when more 
than 30 years ago the discovery of the $J/\psi$ with its small width 
(controlled by $\als$ at the mass scale) acted as an additional 
confirmation  of the QCD asymptotic freedom idea. It is even more the case today
for two reasons. First, remarkable theoretical progress has been achieved 
both in the formulation of nonrelativistic effective field theories
(NR EFTs) for bound states of two heavy quarks \cite{Brambilla:2004jw} 
and in the lattice calculation of nonperturbative matrix elements.
Second,  the last few years have witnessed a kind of New Quarkonium 
Revolution in experiments  with the discovery of  more new states, decays and 
production mechanisms in the last three years  \cite{Brambilla:2004wf,QWG5,Vairo:2006pc} 
than in the entire previous thirty years.

The progress in our understanding of NR EFTs makes it possible to move beyond 
phenomenological models and to provide a systematic description 
inside QCD of all aspects of heavy-quarkonium physics. On the other hand, the recent 
progress in the measurement of several heavy-quarkonium observables makes it meaningful 
to address the problem of their precise theoretical determination.
As we will discuss in the following sections, in this situation heavy quarkonium 
becomes a very special and relevant system to advance our theoretical 
understanding of the strong interactions, also in special environments (e.g. quarkonium in media) 
and in several production mechanisms, as well as  our control of some parameters of the Standard
Model \cite{Brambilla:2004wf,QWG5}.

\section{Theory Developments: Effective Field Theories}
The modern approach to  heavy quarkonium is provided by NR EFTs
\cite{Brambilla:2004jw}. The idea is to
take advantage of the existence of a hierarchy of scales   
to substitute QCD with simpler but equivalent NR EFTs.
A hierarchy of EFTs may be constructed by systematically integrating out 
modes associated to  high energy scales not relevant for the quarkonium system.
Such integration  is made  in a matching procedure that 
enforces the complete equivalence between QCD and the EFT at a given 
order of the expansion in $v$ ($v^2 \sim 0.1$ for $b\bar{b}$,
 $v^2 \sim 0.3$ for $c\bar{c}$,  $v \sim 0.1$ for $t\bar{t}$).
The EFT  realizes a factorization at the Lagrangian level between 
the high energy contributions carried by matching coefficients and 
 the low energy contributions carried by the dynamical degrees of freedom.
The  Poincar\'e symmetry remains  intact at the level of the NR EFT
in a nonlinear realization that imposes exact relations among the 
EFT matching coefficients \cite{poincare0}.

\subsection{Non Relativistic QCD (NRQCD)}
NRQCD is the EFT for two heavy quarks that follows from QCD by integrating out the hard 
scale $m$ \cite{Caswell:1985ui,Bodwin:1994jh}. 
Only the upper (lower) components  of the Dirac fields matter
for quarks (antiquarks) at energies lower than $m$. 

The Lagrangian is organized as an expansion in $v$  and $\als(m)$ of the type:
\be
 {\cal L}_{\rm NRQCD}  = \sum_n  c_n(m,\mu)  \times  O_n/m^n , 
\label{eftnrqcd}
\ee
$\mu$ being the EFT factorization scale. The NRQCD matching  coefficients 
$c_n$ are series in $\als $ and encode the high energy contributions. 
The low energy operators $O_n$  are constructed out of two or four 
heavy quark/antiquark fields plus gluons. The operators bilinear in the fermion 
(or in the antifermion) fields are the same that can be  obtained from a 
Foldy--Wouthuysen transformation of the QCD Lagrangian. Four fermion operators have
to be added. Matrix elements of $O_n$ depend on the scales $\mu$, $mv$, $mv^2$ and $\lQ$.
Hence, operators are  counted in powers of $v$. 
The imaginary part of the coefficients of the 4-fermion operators
contains the information on heavy quarkonium annihilation.
The NRQCD heavy quarkonium Fock state is given by a series of terms, 
increasingly subleading, where the leading term is a  $Q\bar{Q}$ 
in a color singlet state and the first correction, suppressed in $v$,
comes from a $Q\bar{Q}$ in an octet state plus glue.
NRQCD is suitable for studies of spectroscopy (on the lattice),
inclusive decays and production.

\subsection{potential Non Relativistic QCD (pNRQCD)}
In NRQCD, the soft and ultrasoft scales are dynamical. This
results in an ambiguous power counting and in calculations still
complicated by the presence of two scales.
In the last decade, the problem of systematically treating the remaining
dynamical scales in an EFT framework has been addressed  by several groups
\cite{group} and has now reached a good level of understanding.
So one can go down one step further and integrate out also the soft scale,  
matching to the lowest energy EFT that can be introduced for quarkonia, 
where only the ultrasoft degrees of freedom remain dynamical.
Potential Non Relativistic QCD (pNRQCD) \cite{Pineda:1997bj,Brambilla:1999xf,Brambilla:2004jw} 
is the EFT for two heavy quark systems that follows from NRQCD 
by integrating out the soft scale $mv$. The leading order
equation of motion is the Schr\"odinger equation whose potential 
is a matching coefficient of pNRQCD.

Depending on the size of the quarkonium we may distinguish two situations. 
When $mv^2  \simg \lQ$ we speak  about  weakly coupled pNRQCD  because the soft scale is perturbative  
and the matching from NRQCD to pNRQCD may be performed in perturbation theory.
The degrees of freedom are $Q\bar{Q}$ states, singlet and octet in color,
and (ultrasoft) gluons, which are multipole expanded.
The Lagrangian is given by an expansion of the type
\be
\sum_{k,n} {c_k(m, \mu) \over m^k}  \times  V_n(r,
\mu^\prime, \mu)   \times  O_{n,k}\; r^n  ,
\ee
$V_n$ being the pNRQCD matching coefficients. The bulk of the interaction is 
carried by potential-like terms, but non-potential interactions, 
associated with the propagation of low energy degrees of freedom
are present as well and start to contribute at NLO in the multipole expansion. 
They are typically related to nonperturbative effects \cite{Brambilla:1999xf}. 
Matrix elements of $O_{n,k}$ depend on the scales $\mu^\prime$, $mv^2$ and $\lQ$.

When $mv  \sim \lQ$ we speak about  strongly coupled pNRQCD  because the soft scale is nonperturbative  
and the matching from NRQCD to pNRQCD may not be performed in
perturbation theory. The matching coefficients may 
be obtained in the form of expectation values of gauge invariant Wilson loop operators. 
In this case, away from threshold (when heavy-light meson pair and heavy hybrids 
develop a mass gap of order $\lQ$ with respect to the energy of the $Q\bar{Q}$ pair), 
the quarkonium singlet field $ S$ is  the only low energy dynamical 
degree of freedom in the pNRQCD Lagrangian (neglecting pions and other Goldstone bosons), 
which reads \cite{Brambilla:2000gk,Pineda:2000sz,Brambilla:2004jw}:
\be
\quad  {\cal L}_{\rm pNRQCD}=  { S}^\dagger
   \left(i\partial_0-\frac{{\bf p}^2}{2m}-V_S(r)\right){S} . 
\label{sc}
\ee
The potential $V_S(r)$ is a series in the expansion in the inverse of the quark masses;
static, $1/m$ and $1/m^2$ terms have been calculated, see  \cite{Brambilla:2000gk,Pineda:2000sz}.
They involve NRQCD matching coefficients and low energy nonperturbative parts given in terms
of Wilson loops and field strengths insertions in the Wilson loop.
In this regime, from pNRQCD we recover the quark potential singlet model. 
However, here the potentials are calculated from QCD by nonperturbative 
matching. Their evaluation requires calculations on the lattice 
\cite{Bali:2000gf} or in QCD vacuum models \cite{Brambilla:1999ja,Brambilla:1998bt}.

Along the same lines also pNRQCD for $QQ$ states (relevant for doubly charmed
baryons) \cite{Brambilla:2005yk,Fleming:2005pd} and for $QQQ$ states \cite{Brambilla:2005yk}   
has been constructed in the two regimes. Recently  the first lattice calculation of 
the $QQq$ potential has appeared \cite{Yamamoto:2007pf}.

\subsection{Present Reach of Theory}
The physical reach of NRQCD and pNRQCD (combined for some processes  
with soft collinear effective theory, SCET) for heavy quarkonium
includes spectra, inclusive and semi-inclusive decays, transitions and production.

For what concerns spectra and decays, the recent understanding 
of the renormalization group logarithm resummation for correlated scales \cite{Pineda:2001ra} 
and of renormalon subtraction  (for a review see \cite{Brambilla:2004jw})
has impressively extended the reach of QCD  higher order perturbative 
calculations. Moreover, the reduction  in the number of 
nonperturbative matrix elements obtained at the level of pNRQCD 
has greatly enhanced the predictive power of the theory \cite{Brambilla:2004jw}.

Among recent applications, we would like  to recall:
the precise determination of the masses of the $b$ and $c$ quark from quarkonium
with an error better than $50$ MeV (see e.g. the average masses and the errors given in 
\cite{Brambilla:2004wf}, see also \cite{Brambilla:2004jw,charm});
the  recent extraction  of $\als$ from $\Upsilon(1S)$ decay 
resulting in  $\als(M_Z)= 0.119^{+0.006}_{-0.005}$ in agreement 
with the central value of the PDG  and with competitive errors \cite{Brambilla:2007cz};
studies of $t \bar{t}$ production near threshold presently accurate 
at NNLO in perturbation theory with the complete logarithm resummation at 
NLL \cite{Pineda:2006ri,Hoang:2006pc}; a  full understanding  of the  photon 
spectrum of radiative $\Upsilon(1S)$ decays measured by CLEO \cite{GarciaiTormo:2005ch}. 

For the implications of quarkonium on the search for new physics see \cite{Fullana:2007uq}.

In the following, we will briefly summarize the present theoretical status 
for few selected examples.

\section{Potentials and Static Energy}
The $Q\bar{Q}$ potential is a Wilson matching coefficient of pNRQCD 
obtained by integrating out all degrees of freedom but the ultrasoft ones.
If the quarkonium system is small, the soft scale is perturbative and the 
potential can be entirely calculated in perturbation theory 
\cite{Brambilla:2004jw}. It   undergoes renormalization, 
develops a scale dependence and satisfies renormalization
group equations, which eventually allow to resum potentially large logarithms.
The static singlet potential is known at three loops apart from the 
constant term. The first logarithm related to ultrasoft effects arises at three 
loops. Such logarithm at N$^3$LO and the single logarithm at N$^4$LO may be extracted respectively 
from a one-loop and two-loop  calculation in the EFT and have been calculated 
in \cite{Brambilla:1999qa,Brambilla:2006wp}.
The static energy, given by the sum of a constant, the static potential 
and the ultrasoft corrections, is free from renormalon ambiguities. By comparing it
(at NNLL order) with lattice calculations one sees that the QCD perturbative series 
converges very nicely to and agrees with the lattice data in the short range 
and that therefore no large linear (``stringy'') contribution to the static potential exists
at short distances \cite{Pineda:2002se,Brambilla:2004jw}.

\section{Perturbative calculations of Spectra}
In the weak coupling, the soft scale is perturbative  and  the potentials
are purely perturbative objects. Nonperturbative effects enter 
energy levels and decay widths calculations in the form of local or nonlocal 
condensates \cite{Brambilla:1999xj}. We still lack a precise 
and systematic knowledge  of such nonperturbative purely glue 
dependent objects. It would be important to have for them 
lattice determinations or data extraction (see e.g. \cite{Brambilla:2001xy}).
The leading electric and magnetic nonlocal correlators may be related 
to the gluelump masses \cite{Brambilla:1999xf} and to some existing lattice 
(quenched) determinations \cite{Brambilla:2004jw}.  

However, since the nonperturbative contributions  are suppressed in the power 
counting it is possible to obtain good determinations of the masses of the
lowest quarkonium resonances with purely perturbative calculations
in the cases in which the perturbative series converges well 
(i.e. after the appropriate subtractions of renormalons have been
performed and  large logarithms have been resummed).  
In this framework,  power corrections are unambiguously defined.
Renormalon subtraction has been  exploited in \cite{Brambilla:2000db}
to get a prediction of the $B_c$ mass. The NNLO calculation 
with finite charm mass effects \cite{Brambilla:2001qk}
predicts  a mass of $6307(17)$ MeV that matches well the CDF measurement  \cite{Abulencia:2005usa}
and the lattice determination \cite{Allison:2004be}.
The same procedure seems to work at NNLO even for higher states
(inside larger theoretical errors) \cite{Brambilla:2001qk}.
Including logarithm resummation at NLL, it is possible to obtain a 
prediction for the mass of the $\eta_b$, which is $9421 \pm 11^{+9}_{-8}$ MeV  
and for the $B_c$ hyperfine separation, $\Delta=65 \pm 24 ^{+19}_{-16}$ MeV
\cite{Kniehl:2003ap}. A NLO calculation reproduces in part the $1P$ fine splitting 
\cite{Brambilla:2004wu}.

\section{Lattice calculations of Potentials and Spectra}
Traditionally NRQCD lattice calculations have been used to obtain  the  spectrum 
of the low lying $b\bar{b}$ and $c \bar{c}$ states. However the difficulty of the 
calculation of the NRQCD matching coefficients in the lattice scheme 
combined with the problem of the nonperturbative renormalization of the 
zeroth order NRQCD Lagrangian have in part  hampered this approach.
Recent unquenched results exist for $b\bar{b}$
(with tree level matching coefficients) \cite{Gray:2005ur} while for 
the $c\bar{c}$ the current trend is to use unquenched 
relativistic actions and anisotropic lattices \cite{Davies:2006tx}.

In strongly coupled pNRQCD, the energy spectrum is obtained 
by solving the Schr\"odinger equation (\ref{sc}) with the 
potentials  given in terms of the NRQCD matching 
coefficients times expectation values of Wilson loops with field 
strength insertions  to be calculated on the lattice.
Recently the $1/m$ potential and the spin dependent and ``velocity'' dependent 
potentials at order $1/m^2$  have been calculated on the lattice 
with unprecedented precision \cite{Koma:2006si}.
In the long range, the spin-orbit potentials show, for the first 
time, deviations from the flux-tube picture of chromoelectric confinement.
Since a fully consistent renormalization of the EFT operators is still missing 
in the lattice analysis, it may be premature to draw any definitive 
conclusion. However, progress has been made recently in this direction.
In \cite{Guazzini:2007bu}, the nonperturbative  renormalization 
of the chromomagnetic operator in the Heavy Quark Effective Theory, 
which crucially enters in all spin-dependent potentials,
has been performed for the first time. 

The relations among the potentials imposed in pNRQCD by Poincar\'e invariance \cite{poincare0},  
have been checked on the lattice at the few percent level.

The zeroth order pNRQCD Lagrangian is renormalizable. Hence, pNRQCD  
may  be well suited for direct lattice evaluation of quarkonium correlation functions.

\section{Quarkonium Decays and Transitions}
Expressions for inclusive electromagnetic and hadronic quarkonium decays 
are now known at order $v^7$ in NRQCD \cite{Bodwin:2002hg,Brambilla:2006ph}.
The matching coefficients are known at different accuracy 
in the $\als$ expansion, for a review  see \cite{Vairo:2003gh}. 
At the moment, specific problems for phenomenological applications 
arise from  the proliferation in the number of unknown nonperturbative matrix elements 
in NRQCD and the bad convergence of the perturbative series of some NRQCD matching 
coefficients. Only few NRQCD matrix elements have been calculated on the lattice up to now 
(see e.g. \cite{Bodwin:2005gg}). A significant reduction  in the number of 
nonperturbative  operators for inclusive decays is achieved in strongly coupled pNRQCD, where 
the  NRQCD decay matrix elements factorize in a part, which is the wave function in
the origin squared (or its derivatives), and in a part which contains 
gluon tensor-field correlators \cite{Brambilla:2001xy,Brambilla:2002nu,Brambilla:2003mu}.

For the lowest resonances, inclusive decay widths are given in weakly coupled pNRQCD by a 
convolution of perturbative corrections and nonlocal nonperturbative correlators.
The perturbative calculation embodies large contributions and 
requires the resummation of large logarithms (see e.g. A. Pineda in \cite{QWG5}).
Recently, higher order contributions to quarkonium production
and annihilation have been obtained \cite{Beneke:2007pj}.

Allowed magnetic dipole transitions between $c\bar{c}$ and $b \bar{b}$  ground states
have been considered in pNRQCD at NNLO  in \cite{Brambilla:2005zw}.
The results are: $\Gamma(J/\psi \to \gamma \, \eta_c) \! = (1.5 \pm 1.0)~\hbox{keV}$
and $\Gamma(\Upsilon(1S) \to \gamma\,\eta_b)$ $=$  $(k_\gamma/39$ $\hbox{MeV})^3$
$\,(2.50 \pm 0.25)$ $\hbox{eV}$, where  the errors account for uncertainties 
coming from higher-order corrections. The width $\Gamma(J/\psi \to \gamma\,\eta_c)$ 
is consistent with the PDG value. Concerning $\Gamma(\Upsilon(1S) \to \gamma\,\eta_b)$, 
a photon energy $k_\gamma = 39$ MeV corresponds to a $\eta_b$ mass of 9421 MeV. 
The pNRQCD calculation features a small quarkonium magnetic moment (in agreement 
with a recent lattice calculation \cite{Dudek:2006ej}) and the interesting 
fact,  related to the Poincar\'e invariance of the NR
EFT, that M1 transition of the lowest quarkonium states at relative order $v^2$ are 
completely accessible in perturbation theory \cite{Brambilla:2005zw}.

\section{Quarkonium Production}
Although a formal proof of the NRQCD factorization formula for heavy quarkonium production 
has not yet been obtained, NRQCD factorization has proved to be successful 
to explain a  variety of quarkonium production processes (for a review see e. g. 
the production chapter in \cite{Brambilla:2004wf} and the more 
recent review \cite{Lansberg:2006dh}).
In the last years, there has been progress toward an all order proof.
In \cite{Nayak:2006fm}, it has been shown that a necessary condition 
for factorization to hold at NNLO is that the conventional octet NRQCD production matrix elements 
must be redefined by incorporating Wilson lines that make them manifestly gauge 
invariant. Differently from decay processes, a pNRQCD treatment does not exist so far for 
quarkonium production. In the last years, two main problems have plagued our understanding 
of heavy quarkonium production. The first  BELLE and BABAR measurements of the cross section 
$\sigma(e^+e^-\to J/\psi+\eta_c)$ were about one order of magnitude above theoretical expectations.
Triggered by this, some errors have been corrected in some of the theoretical 
determinations, and, more relevant, NLO corrections in $\als$ 
and in $v^2$ have been calculated and some class of relativistic 
corrections has been resummed \cite{Zhang:2005cha,Zhang:2006ay,Bodwin:2006ke,
Bodwin:2006dm,Bodwin:2007ga}. For calculations in the framework of  light cone 
see \cite{Braguta:2005kr}.\par
One can say that 
now the  discrepancy between the QCD theoretical prediction for 
$\sigma(e^+e^- \to J/\psi+\eta_c)$ and the experimental measurements has been resolved.
However the discrepancy seems to survive for the inclusive production 
$\sigma(e^+e^- \to J/\psi+ c\bar{c})$ where relativistic corrections are tiny \cite{He:2007te}.
In addition,  the latest data on  charmonium and bottomonium polarization at Tevatron (Run II)
\cite{QWG5}  contradict the prediction of NRQCD with traditional power counting. 
Recently, singlet contributions to quarkonium hadroproduction have been calculated at NLO
\cite{Campbell:2007ws}  and hadroproduction  of heavy quarkonium in association with an additional
heavy quark pair has been calculated at LO \cite{Artoisenet:2007xi}. Both contributions turn out to 
be seizable and  tend to unpolarize the produced quarkonium.
A  modification 
of the NRQCD factorization approach for processes involving production in association
with another heavy quark pair may be necessary \cite{Nayak:2007mb}.

\section{Theory Open Challenges}

\subsection{Threshold States}
For states near or above threshold a general systematic theoretical treatment
has still to be developed. At the moment, the first preliminary studies 
of excited resonances on the lattice are just appearing \cite{Juge:2006fm,Ehmann:2007hj,Dudek:2007wv}
some of them being still quenched. Most of the existing analyses have 
therefore to rely on phenomenological models. 

However, in some cases, a theoretical treatment based on an EFT approach has been developed. 
This is notably the case of the $X(3872)$ in the molecular picture \cite{Braaten:2003he}.
Most of the newly discovered states in the charmonium region 
lie close to threshold or over threshold. The confirmation 
of some of these new states would require a trustable calculation of individual 
contributions and interference terms in the total cross section. 
{\it It  is high priority for theory to develop a systematical effective field theory 
approach to quarkonium states close to threshold and coupled  to heavy-light mesons}.

\subsection{Quarkonium at Finite $T$}
Quarkonium suppression is believed to be a clean signal for quark gluon plasma 
formation in heavy ion collisions, to be, however, considered together with
possible  quarkonium  recombination effects in the medium. 
An extensive literature in the field deals both with lattice calculations 
of the free energy of a quark-antiquark pair as well as with model calculations of the 
quark-antiquark correlators and  spectral functions at finite $T$ \cite{Mocsy:2007jz}.
As a matter of fact, it has not yet been understood how to define the quark-antiquark 
static potential at finite $T$, even if recently some steps forward  have been accomplished, obtaining 
a (complex-valued) potential from a perturbative calculation of the Wilson loop \cite{Laine:2006ns}.
{\it It is  a high priority for theory to develop an EFT systematical approach to quarkonium 
physics at finite $T$}, where a potential may be clearly defined and calculations 
may be performed that include all the relevant low energy dynamical degrees of freedom.

\end{document}